\documentclass{aa}  

\usepackage{duckuments} 
\usepackage{graphicx}
\usepackage{xcolor}  
\usepackage{txfonts}
\usepackage{lipsum}
\usepackage{lscape}
\usepackage{subcaption}
\usepackage{lscape}           
\usepackage{placeins}          
\usepackage{natbib}
\usepackage{ulem}
\bibpunct{(}{)}{;}{a}{}{,} 

\def\apjl{{ApJL}}

\begin{document}

   \title{Hubble tension and small-scale inhomogeneities on light propagation}

   \author{L. Kraiselburd\inst{1,2}\corrauth{lkrai@fcaglp.fcaglp.unlp.edu.ar}        
        \and C. Pigozzo\inst{3}\email{cpigozzo@ufba.br}
        \and S.J. Landau\inst{4,5}\email{slandau@df.uba.ar}
        \and J. Alcaniz\inst{6}\email{alcaniz@on.br}
        }

   \institute{Facultad de Ciencias Astronómicas y Geofísicas, Universidad Nacional de La Plata, Paseo del Bosque, B1900FWA, La Plata, Argentina
   \and Consejo Nacional de Investigaciones Científicas y Técnicas (CONICET),Godoy Cruz 2290, Buenos Aires, 1425, Argentina
   \and Instituto de Física, Universidade Federal da Bahia, Salvador, 40210-340, Bahia, Brazil
   \and Universidad de Buenos Aires, Facultad de Ciencias Exactas y Naturales, Departamento de Física, Ciudad Universitaria 1428, Buenos Aires,1460, Argentina
   \and CONICET - Universidad de Buenos Aires, Instituto de Física de Buenos Aires (IFIBA), Ciudad Universitaria 1428, Buenos Aires, 1460, Argentina
   \and Observatorio Nacional, Rio de Janeiro, 20921-400, RJ, Brazil  
   }

   \date{Received XXX 2xx, 20XX}

  \abstract
   {A major observational challenge within the standard cosmological framework is the Hubble tension, a statistically significant ($\sim 5 \sigma$) disagreement between the Hubble constant derived from cosmic microwave background measurements and the value obtained through local distance-ladder methods based on Type Ia supernovae and Cepheid variable stars.}
   {We relax the assumption of the Friedmann-Lemaître-Robertson-Walker (FLRW) distance- redshift relation and explore theinfluence of small-scale inhomogeneities on the propagation of light from distant sources, using the Zeldovich-Kantowski-Dyer-Roeder (ZKDR) approximation as an alternative approach to address this tension.} 
   {We employ two distinct formulations of the ZKDR equation  to test our hypothesis using recent Type Ia supernovae datafrom the Pantheon+ compilation and the SH0ES collaboration and six gravitational lens systems from the H0LiCOW collaboration.}
   {We obtained constraints on the cosmological parameters and the ZKDR model parameters within the framework of the inhomogeneous models considered. The model comparison criterion indicates that the data show weak preference of  $\Lambda$CDM over the  flat ZKDR model , whereas the remaining models studied are strongly disfavored.}
  {Our findings indicate that a background model characterized by the ZKDR approximation and its modifications does not solve or alleviate the Hubble tension.}


  \maketitle

\nolinenumbers

\section{Introduction}

\label{introduction}

Advancements in our understanding of systematic errors, combined with the increased quantity and precision of cosmological data over the past 20 years, have resulted in a more accurate determination of cosmological parameters. Although the standard $\Lambda$-Cold Dark
Matter ($\Lambda$CDM) model can explain most current datasets, there are significant discrepancies in the values of cosmological parameters derived from different data sources within this model\footnote{Measurements of Baryon Acoustic Oscillations from the DESI collaboration \citep{DESI:2025zgx}, combined with data from the cosmic microwave background and Type Ia
supernovae data, have challenged the $\Lambda$CDM paradigm indicating a potential
evolution in the dark energy equation of state. These results are currently the subject of debate \citep{Efstathiou:2024xcq}, with both parametric and non-parametric analyses yielding divergent conclusions \citep{DESI:2024mwx,Dinda:2024ktd,Sousa-Neto:2025gpj,DESI:2025fii}.}.

The most significant issue is known as the Hubble tension, which refers to a discrepancy between the value of the Hubble constant $H_0$ obtained from Cosmic Microwave Background (CMB) data within the $\Lambda$CDM model~\citep{Aghanim:2018eyx} and the value derived from type Ia supernovae (SNIa) and Cepheid variables observations~\citep{2022ApJ...938..110B,2022ApJ...938..113S}. Quantitatively, fitting the $\Lambda$CDM model to the Planck data, we find 
\begin{equation} \label{H0CMB}
H_0 = 67.43 \pm 0.49\ {\rm{km s^{-1}Mpc^{-1}}}\,.    
\end{equation}

In comparison, the value of the Hubble constant measured by the SH0ES collaboration based on Cepheid variables and SNIa observations is 
\begin{equation}
H_0 = 73.01 \pm 0.99\ {\rm{km s^{-1}Mpc^{-1}}}\, ,
\end{equation}
which differs from (\ref{H0CMB}) by more than $5\sigma$. 

High-resolution ground-based experiments~\citep{Aiola_2020,PhysRevD.108.023510} yield independent $H_0$ estimates within the $\Lambda$CDM framework that are consistent with the Planck value, while JWST observations of Cepheids, the tip of the red giant branch, and carbon-rich asymptotic giant branch stars furnish~\citep{2020ApJ...891...57F}
\begin{equation}
H_0 = 69.8 \pm 1.9\ {\rm{km s^{-1}Mpc^{-1}}}\,,    
\end{equation}
which is $\sim 1.5\sigma$ and $\sim 1.2\sigma$ away from the SH0ES and CMB values, respectively (for recent reviews on the $H_0$ tension, we refer the reader to \citet{2021ApJ...919...16F,DiValentino:2021izs,Efstathiou:2024dvn}).

The origin of this tension has sparked considerable debate within the cosmological community. Some analyses argue that systematic errors in the SH0ES data may not have been fully accounted for \citep{Efstathiou:2021ocp,2024arXiv240806153F,2024arXiv240811031P} while others conclude that the $\Lambda$CDM model may be missing new physics and investigate alternative cosmological models (see e.g. \cite{Karwal:2016vyq,Alcaniz:2019kah,Poulin:2018cxd,Alcaniz:2022oow,2024JCAP...04..059K,daCosta:2023mow} and references therein). 

In this paper, we take a different approach to investigate the Hubble tension and explore the global effects of small-scale inhomogeneities in light propagation, while still assuming that the universe is homogeneous and isotropic. This idea was initially explored by Zeldovich, Dashevskii, and Kantowski in their respective studies~\citep{1964SvA.....8...13Z,1965SvA.....8..854D,1969ApJ...155...89K}. It maintains the Friedmann-Lama\^itre-Robertson-Walker (FLRW) background geometry and expansion history but separates matter into two components: one that is smoothly distributed, accounting for a fraction $\alpha$ of the total density, and the other, comprising $1 - \alpha$, which consists of clumps (for a recent review, see \citep{Helbig:2019jcm}). In what follows, we consider the Zeldovich-Kantowski-Dyer-Roeder (ZKDR) distance relation and a modified version of it (mZKDR) to describe the propagation of light rays\footnote{Most studies refer to the distance relation incorporating these concepts as the Dyer-Roeder approximation. Here, we follow \cite{2004IJMPD..13.1309A} and refer to it (see Eq.~\ref{ec:DR}) as the Zeldovich-Kantowski-Dyer-Roeder (ZKDR) distance relation to recognize the contributions of the original authors on this topic.}. We examine both flat and curved universes and consider the possibility that the smoothness parameter of the mZKDR equation varies with redshift. We test these scenarios with SNIa data from the Pantheon+ compilation, as well as low-redshift SNIa data calibrated with Cepheids from the SH0ES collaboration  \citep{2022ApJ...938..110B,2022ApJ...938..113S}. Additionally, we incorporate in our analyses the time delays of gravitational lenses reported by the H0LiCOW collaboration \citep{2020MNRAS.498.1420W}. It is well known that the low sample size of the H0LiCOW data set leads to weak constraints on $H_0$ which also remain consistent with both CMB and SH0ES estimates.  Consequently, addressing the $H_0$ tension in this study requires that the $H_0$ values inferred from type Ia supernovae data, using the modified Dyer-Roeder distances, be consistent with the CMB estimate.

The structure of our paper is as follows: In Section \ref{theory}, we summarize the fundamental principles of the ZKDR and mZKDR equations, detailing how each framework modifies the angular diameter distance. Section \ref{data} provides a brief overview of the data sets used in our analysis, while Section \ref{results} presents and discusses the results of our statistical analysis. Finally, we present our main conclusions in Section \ref{conclusions}.

\section{The ZKDR approximation}
\label{theory}

We first recall the optical scalar equation in the geometric optics approximation \citep{1992grle.book.....S}:
\begin{equation}\label{Da}
    \frac{d^2 \sqrt{A}}{ds^2} +  \tfrac{1}{2} R_{\alpha \beta} k^\alpha k^\beta \sqrt{A} = 0\;,
\end{equation}
where we neglect the optical shear. Here $A$ refers to  the beam cross section area, $s$ is an affine parameter describing the null geodesics, $k^\alpha$ is the tangent vector to the surface of propagation of the light ray and $R_{\alpha \beta}$ is the Ricci tensor. If we assume a universe with presureless matter and a cosmological constant, in comoving and synchronous coordinates, then $R_{\alpha \beta} k^\alpha k^\beta=\kappa \rho_m k^0 k^0$. As mentioned earlier, the key assumption of the ZKDR approximation is that a mass fraction $\alpha$ of the total matter in the Universe is smoothly distributed while a fraction $1-\alpha$ is bound in galaxies. In other words, $\alpha$ represents the fraction of matter homogeneously distributed within the beam. This parameter takes values between 0 and 1 with distinct physical interpretations: when it equals 0, all matter exists in clustered form (empty beam condition), whereas a value of 1 indicates a completely homogeneous matter distribution (filled beam condition).
Noting that the angular diameter distance $D_A$ is proportional to $\sqrt{A}$, Eq. \ref{Da} turns into the ZKDR equation:
\begin{align}\label{ec:DR}
\frac{d^2 \widetilde{D}_A}{dz^2} + \left(\frac{d\ln H}{dz}+\frac{2}{1+z}\right)\frac{d\widetilde{D}_A}{dz} = 
-\frac{3}{2}\Omega_m\frac{H_0^2}{H^2}(1+z)\alpha(z)\widetilde{D}_A\;,
\end{align}
where $\widetilde{D}_A=H_0 D_A/c$ is a dimensionless quantity, $H(z)^2 = H_0^2[\Omega_m (1+z)^3+\Omega_{\Lambda}+\Omega_k(1+z)^2]$, $\Omega_m$, $\Omega_\Lambda$ and $\Omega_k$ are the matter (dark + baryonic), dark energy and curvature parameters, respectively,  and the smoothness parameter $\alpha (z)$ can be constant or a function of $z$. Thus, our first expression for the light propagation in such a background will be Eq. \ref{ec:DR}, which reduces to the usual FLRW $\Lambda$CDM distance-redshift relation for $\alpha = 1$. 

We note that Eq. \ref{ec:DR}  retains the form of the smooth FLRW background, with $\rho_m$ replaced by $\alpha \rho_m$ only on the RHS to account for the effects of inhomogeneities. In this way variations in $\rho$ along any null geodesic are compensated by corresponding fluctuations in shear and curvature, a scenario that appears physically implausible.  To avoid this problem \cite{2012MNRAS.426.1121C} proposed a modified version of the ZKDR distance-relation.
Starting from the FLRW expression for $D_A$, they replaced $\tfrac{dH}{dz}$ for the FLRW expression and, replaced $\rho_m$ by $\alpha \rho_m$ everywhere.

The modified formula (revised by \cite{2021MNRAS.503.3179K}) is given by:
\begin{align}\label{ec:mDR}
\frac{d^2 \widetilde{D}_A}{dz^2} +& \left(\frac{(1+z)H_0^2}{2H^2}[3\alpha(z)\Omega_m (1+z)+2\Omega_k]+\frac{2}{1+z}\right)\frac{d\widetilde{D}_A}{dz} = \nonumber \\ 
& -\frac{3}{2}\Omega_m\frac{H_0^2}{H^2}(1+z)\alpha(z)\widetilde{D}_A,
\end{align}
where $H(z)^2 = H_0^2[\alpha(z)  \Omega_m (1+z)^3+\Omega_{\Lambda}+\Omega_k(1+z)^2]$, and the smoothness parameter $\alpha (z)$ has now been included in all density terms ($\rho_m \rightarrow \alpha\rho_m$), but its derivatives have been neglected. Thus, Eq. (\ref{ec:mDR}) describes changes in the expansion dynamics caused by local inhomogeneities. This is a more accurate attempt to model global effects of small-scale inhomogeneities in light propagation, and we will refer to it as mZKDR model. We recall that the initial conditions to solve Eqs. (\ref{ec:DR}) and (\ref{ec:mDR}) are $\widetilde{D}_A(z=0) =0$ and ${d\widetilde{D}_A}/{dz}|_{z=0}=1$. We note that in Eq.\ref{ec:mDR} $\alpha$ and $\Omega_m$ are correlated while this degeneracy is broken in Eq. \ref{ec:DR}. Therefore, we expect that the data will provide tighter constraints on $\alpha$ in  ZDKR model compared to the mZKDR case.

On the other hand, several authors \citep{2006astro.ph..9129S,2011MNRAS.412.1937B,2021MNRAS.503.3179K,2012MNRAS.426.1121C} have considered the possibility that the unbounded matter fraction $\alpha$ is a function of the redshift. To compare with the observational data set described in Section \ref{data}, we consider different behaviors for the smoothness parameter proposed in the literature. Table \ref{tab:models} shows  specific parameterizations of $\alpha(z)$, the corresponding reference, and the label we adopt to report the results in Section \ref{results}. The parameters $\alpha_0$, $\alpha_1$, $\beta_0$ and $\gamma$ are constants. In this work, their values will be estimated using data from SNIa and gravitational lenses.

The tension between CMB data and the latest BAO data from the DESI DR2 release can be alleviated by considering a non-flat universe \citep{2025arXiv250500659C,2025arXiv250415190D}.  Furthermore, the CMB data reported by the Planck collaboration \citep{Aghanim:2018eyx} are compatible with an open universe exhibiting a small curvature. Therefore, we will also analyze the scenario of a non-flat universe with the light propagation described by the Dyer-Roeder equation.

An important clarification is that  the inhomogeneities modeled by the smoothness parameter $\alpha$, are not incorporated as a source for the metric perturbations. The latter are typically defined through a linear expansion of the metric tensor around the FLRW background, whereas the smoothness parameter is an empirical quantity introduced to account for the cumulative effect of numerous unresolved small-scale structures (such as clumps of dust) along the path of a light ray. 

\begin{table}[h]
    \centering
    \footnotesize
    \caption{Parameterizations of the smoothness parameter $\alpha$.}
    \label{tab:models}
    \begin{tabular}{cc}
        \hline
    \hline
       Parameterization  & $\alpha(z)$  \\
       \hline
        mZKDR& $\alpha$ \tablefootmark{(a)}   \\
       mZKDR1& $\alpha_0+\alpha_1 z$ \tablefootmark{(b)} \\
       mZKDR2 & $\frac{\beta_0(1+z)^{3\gamma}}{1+\beta_0(1+z)^{3\gamma}}$ \tablefootmark{(c)} \\
   \hline
    \hline
    \end{tabular}
    \tablefoot{The first column indicates the label for each parameterization as reported in Sect. \ref{results}.\\
\tablefoottext{a}{ Parameterization from Kalomenopoulos et al. (2021).}
\tablefoottext{b}{Parameterization  from Linder (1988).}
\tablefoottext{c}{Parameterization  from Santos \& Lima (2006). }}
\end{table}

\begin{table}[h]
\footnotesize
\centering
\caption{Priors used for each model in the statistical analyses.
}
\label{Tab: prior1}
\begin{tabular}{c c c c  }
\hline
Model&  $\alpha$ &         &  $\Omega_k$    \\
\hline 
ZKDR &  $[0,1]  $       & ---                                         & 
---    \\
mZKDR &   $[0,1]  $    &--- & 
---   \\
 $\Lambda$CDM                                               &   ---   &---  &   
---   \\
non-flat ZKDR  &   $[0,1]  $ &---   &   
$[-0.5,1]$  \\
non-flat mZKDR  &   $[0,1]  $  &---  &  
$[-0.5,1]$    \\
 non-flat $\Lambda$CDM                                               &   ---     &  ---& 
$[-0.5,1]$\\
mZKDR1 &  $\alpha_0=$$[0,1]  $    & $\alpha_1=$  $[0,1]  $& ---                \\
mZKDR2 &   $\beta_0=$$[0,1]  $    & $\gamma=$  $[0,1]  $   & ---              \\
\hline \hline
\end{tabular}
\tablefoot{$\Omega_m$, $H_0$ and $M_{abs}$ priors in all analyzed cases are given by $[0.05,1]$, $[60,80]$  and $[-19.60 , -19.10]$ respectively}.
\end{table}

\section{Data sets} \label{data}

In this section, we describe the data sets we use to constrain the inhomogeneous cosmological models described by the Dyer-Roeder equation.  We use type Ia supernovae (SNIa) data and the time delays of strong lensed quasars from the HOliCOW collaboration. The time delays  depend directly on the angular diameter distances to the source and lens, which is the quantity governed by the Dyer-Roeder equation. On the other hand, the luminosity distance, which is related to the SNIa observable ,  has also a simple relation with the diameter angular distance. We will show in this section that both data sets are in agreement, in the sense that the confidence intervals inferred with each data set separately are consistent, which is the necessary condition to perform the joint statistical analysis.

\subsection{Type Ia supernovae}
The homogeneity of type Ia supernovae (SNIa) spectral and light curves makes them ideal observational objects for determining distances and constraining cosmological parameters. Additionally, the vast amount of data collected in all directions strengthens this conclusion. The distance modulus $\mu$ can be obtained from the SNIa light curves,
\begin{equation}
\mu=m_b-M_{abs}
\label{mu_obs}
\end{equation}
where $m_b$  is an overall flux normalization and $M_{abs}$ the absolute magnitude of the star; and from the following theoretical expression
\begin{equation}
\mu=25+\log_{10}\left[d_L(z)\right]\;,
\end{equation}
being $d_L(z)=(1+z)^2D_A(z)$ the luminosity distance.

We will consider two supernovae data sets  \footnote{Each data set contains the B band magnitude from
SNIa and its corresponding redshift together with the corresponding distance
modules from Cepheid data.}: i)one comprising the calibrator data selected by the SH0ES collaboration \citep{2022ApJ...934L...7R} and a subset of the Pantheon$^+$ compilation  and ii) another that includes the complete Pantheon$^+$ data set \citep{2022ApJ...938..110B,2022ApJ...938..113S} 
We will refer to the first set as SH0ES+HF (SH0ES+Hubble flow) and to the second as PPS (Pantheon$^+$+SH0ES). The SH0ES+HF data set includes 77 data points that belong to 42 SNIa at redshift $z<0.01$ which are called calibrators (SH0ES). In addition, this data set also includes 277 Hubble flow SNIa from Pantheon$^+$ within a redshift range $0.023 < z < 0.15$ (HF). The latter are hosted in late-type galaxies like the Cepheids. 
The PPS data set corresponds to the full Pantheon$^+$ compilation, consisting of 1657 data points spanning the redshift range $0.01 < z < 2.26$ \footnote{The entire compilation consist of 1701 data points. However, for the purpose of testing cosmological models only the points at redshift $z>0.01$ are considered.}.  This compilation also incorporates the SH0ES data set ($z<0.01)$ for the purpose of calibration.
The likelihood of the SNIa data reads.
\begin{equation}
\ln {\cal L} = - \dfrac{1}{2} \left( \Delta\mu^{T} \cdot C^{-1} \cdot \Delta\mu \right) \, ,
\end{equation}

where $\Delta\mu=\mu^{th}-\mu^{obs}$ is a vector that contains the difference between the theoretical and observational distance modulus of all measurements in the compilation,
\begin{align}
\Delta\mu&= \begin{cases} 
m_{b}-M_{abs} -  \mu^{th}(z) & \text{if Hubble diagram supernova,} \\
m_b-M_{abs} -  \mu_{{\rm ceph}} & \text{if supernova in Cepheid host,}  
\end{cases} 
\end{align}
and $C=\Sigma_\text{sne}+\Sigma_\text{ceph}$ is the covariance matrix reported in \cite{2022ApJ...938..113S} that includes correlations between data. For calibrators, the theoretical distance modulus is replaced with the distance modulus obtained from Cepheids $\mu_{\rm ceph}$.

Equation \ref{mu_obs} is a simplification of the Tripp formula \citep{Tripp1998}, where the corrections to the distance modulus are already included and the {\it nuisance parameters} are determined assuming a given scenario~\citep{Negrelli2020,Leizerovich2022}. 

Moreover, the relative magnitudes of SNIa ($m_b$) are derived quantities obtained after light-curve fitting, calibration, and several observational corrections. Their corresponding covariance matrix includes multiple statistical and systematic contributions  provided by the PPS collaboration. In addition, the high signal-to-noise ratio of the SNIa data justifies the Gaussian approximation through the central limit theorem. Therefore, a multivariate Gaussian likelihood is the standard approach in SNIa cosmological analyses, although photon counts are intrinsically Poisson distributed\footnote{Recent studies \cite{2024JHEAp..41...30D,2025MNRAS.536..234L}, have explored alternative likelihoods for the Pantheon+ dataset, including a Student's $t$-distribution. However, the preferred values of the degrees of freedom are sufficiently large that the distribution is nearly Gaussian, yielding only minor effects on cosmological parameter estimates and the Hubble tension. We repeated our analysis using this alternative likelihood and found slightly tighter constraints only for the $\Lambda$CDM model (for the ZKDR and mZKDR models, the Gaussian distribution is more constraining), while our main conclusions remained unchanged. Therefore, for consistency with the standard methodology adopted in most SNIa cosmological studies, we present our results assuming a Gaussian likelihood.}.

\subsection{Gravitational lenses (H0LiCOW)}

The phenomenon of gravitational lensing illustrates that the assumption of a completely homogeneous universe cannot accurately describe light propagation. Gravitational lensing occurs when light rays from distant, bright objects are bent by the presence of a massive object (acting as a lens) located between the emitting and receiving objects, potentially generating multiple images of the same source. Since the travel time of light from the source to the observer depends on both the length of the path and the gravitational potential it traverses along the way, those rays that pass through a lens experience a delay in time compared to those that do not. The delays in time of two images ({\it i} and {\it j}) generated by the same source through a plane lens can be expressed as \citep{1992grle.book.....S}
\begin{equation}
\Delta t_{ij} = \frac{D_{\Delta t}}{c} \left[\frac{(\boldsymbol{\theta}_i - \boldsymbol{\beta})^2}{2} - \psi(\boldsymbol{\theta}_i) - \frac{(\boldsymbol{\theta}_j - \boldsymbol{\beta})^2}{2} + \psi(\boldsymbol{\theta}_j)\right]\,,
\end{equation}
where $\boldsymbol{\theta}_{i/j}$ and $\psi(\boldsymbol{\theta}_{i/j})$ represent the angular position and  the lens potential at the image position of each image, and $\boldsymbol{\beta}$ the source position. Meanwhile, $D_{\Delta t}$ is the time-delay distance \citep{Refsdal:1964nw,1992grle.book.....S,Suyu:2009by} given by,
\begin{equation}
D_{\Delta t} \equiv (1 + z_d) \frac{D_{A_{d}} D_{A_{s}}}{D_{A_{ds}}}\,,
\end{equation}
with $z_d$ representing the lens redshift. $D_{A_{d}}$, $D_{A_{s}}$, and $D_{A_{ds}}$ refer to the angular diameter distances to the lens, to the source, and between the lens and source, respectively. 
The time delay $\Delta t_{ij}$ is measured from the  exhaustive tracking of images fluxes, and both the potentials and the source position are determined by a mass model of the system. In this work, we use six lens systems released by the H0LiCOW compilation \citep{2020MNRAS.498.1420W}: B1608+656, RXJ1131-1231, HE 0435-1223, SDSS 1206+4332, WFI2033-4723 and PG 1115+080, within a source redshift $0.65<z_s<1.789$. We used the likelihood provided by the H0LiCOW collaboration \footnote{https://shsuyu.github.io/H0LiCOW/site/}. According to \cite{2020MNRAS.498.1420W}, the time delay measurements for each lens are uncorrelated. Furthermore, the authors compute the Bayes factor between different lenses to confirm the consistency of all measurements.

\section{Results and discussion}\label{results}
\begin{figure*}
\begin{subfigure}{.5\textwidth}
  \centering
  \includegraphics[width=.8\linewidth]{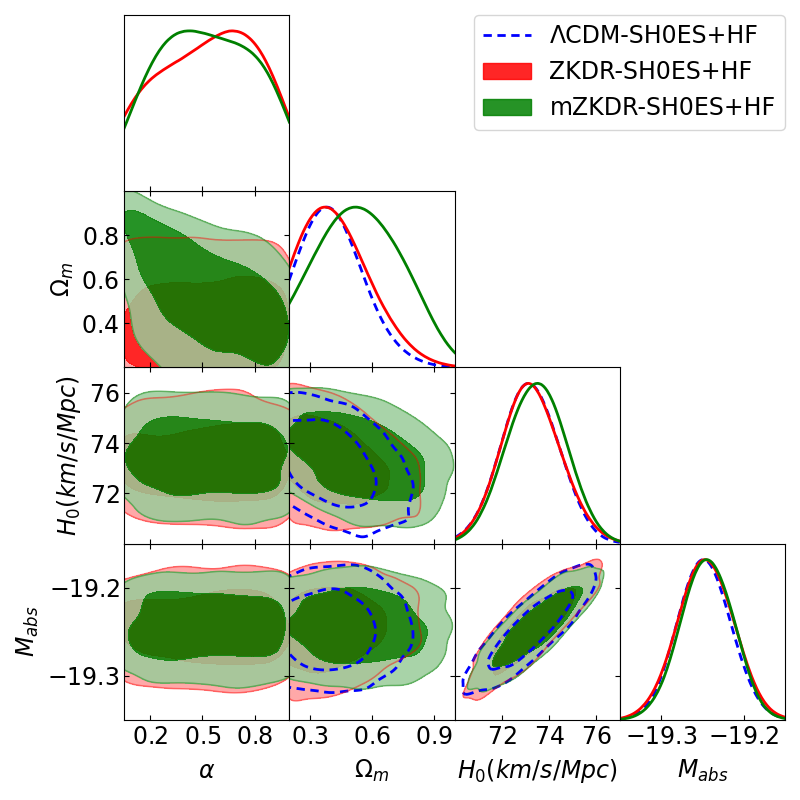}
  \caption{}
  \label{fig:sfig01}
\end{subfigure}%
\begin{subfigure}{.5\textwidth}
  \centering
  \includegraphics[width=.8\linewidth]{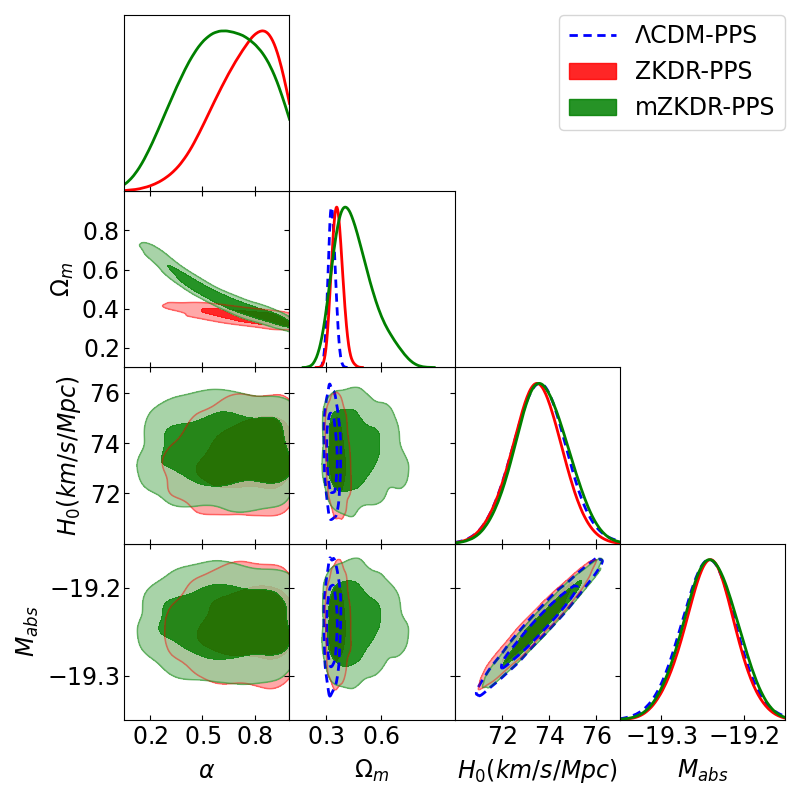}
  \caption{}
  \label{fig:sfig02}
\end{subfigure}
\caption{Results of the statistical analyses assuming a flat universe and constant $\alpha$. The darker and brighter regions correspond to 65\% and 95\% confidence levels, respectively. The plots in the diagonal show the posterior probability density for each of the free parameters of the scenarios.  The left panel shows the results for SH0ES+HF dataset only while the right panel shows the results for PPS. For comparison, the dashed blue curves represent the analyses for the standard (FLRW) $\Lambda$CDM model ($\alpha=1$). }
\label{fig:fig0}
\end{figure*}

\begin{figure*}
\begin{subfigure}{.5\textwidth}
  \centering
  \includegraphics[width=.8\linewidth]{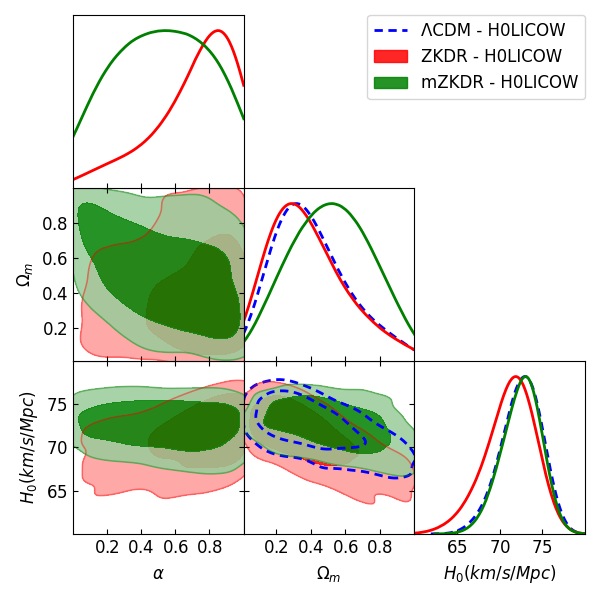}
  \caption{}
  \label{fig:sfig1}
\end{subfigure}
\begin{subfigure}{.5\textwidth}
  \centering
  \includegraphics[width=.8\linewidth]{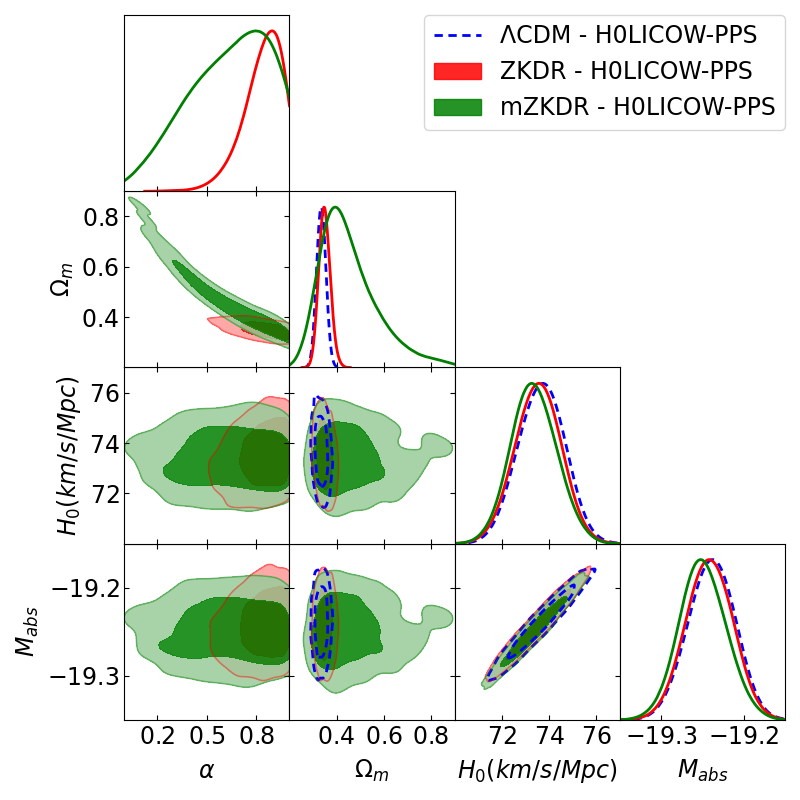}
  \caption{}
  \label{fig:sfig2}
\end{subfigure}
\caption{Results of the statistical analyses assuming a flat universe and constant $\alpha$. The darker and brighter regions correspond to 65\% and 95\% confidence levels, respectively. The plots in the diagonal show the posterior probability density for each of the free parameters of the scenarios.  The left panel shows the results for H0LiCOW data only while the right panel shows the results for both H0LiCOW and PPS data. For comparison, the dashed blue curves represent the analyses for the standard (FLRW) $\Lambda$CDM model ($\alpha=1$). }
\label{fig:fig1}
\end{figure*}

\begin{figure*}
\begin{subfigure}{.5\textwidth}
  \centering
  \includegraphics[width=.8\linewidth]{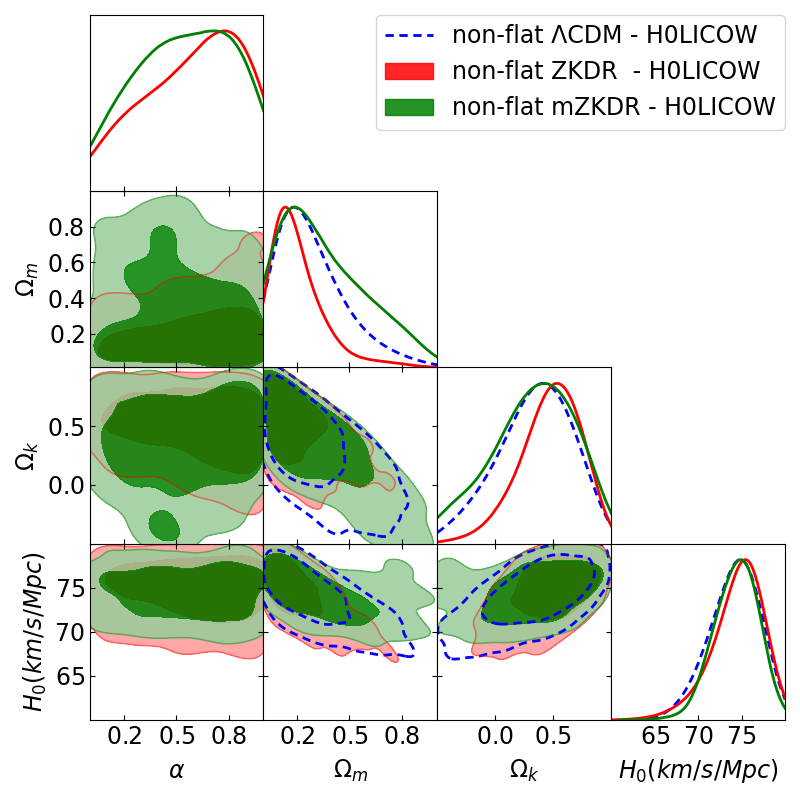}
  \caption{}
  \label{fig:sfig3}
\end{subfigure}%
\begin{subfigure}{.5\textwidth}
  \centering
  \includegraphics[width=.8\linewidth]{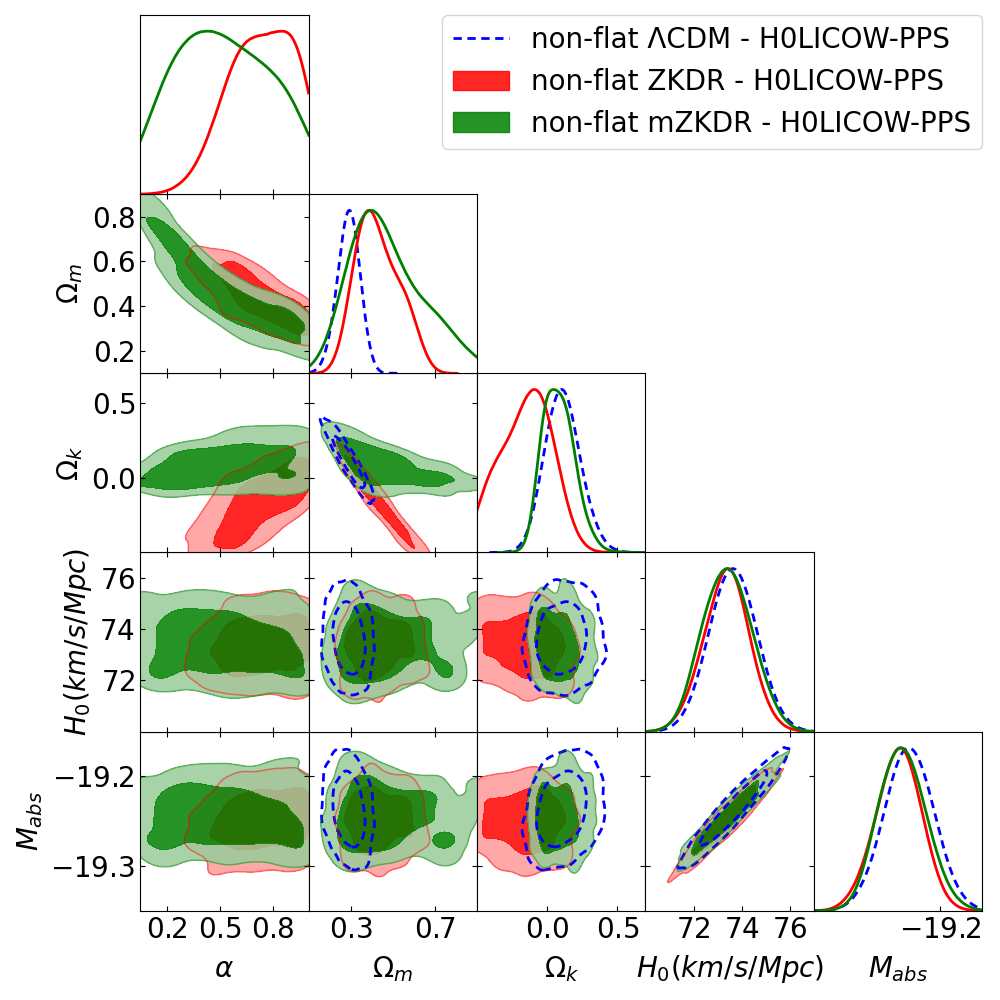}
  \caption{}
  \label{fig:sfig4}
\end{subfigure}
\caption{The same as in previous figure for non-flat geometries.}
\label{fig:fig2}
\end{figure*}

In this section, we show the results of our statistical analyses assuming the ZKDR and mZKDR distance relations presented in Section \ref{theory} and the observational data described in Section \ref{data}. For comparison, we also show the results for the standard (FLRW) $\Lambda$CDM model. The free parameters in our analysis are: the smoothness parameter $\alpha$, the mass density parameter $\Omega_m$, the Hubble parameter $H_0$, the curvature parameter $\Omega_k$ -- in those cases where a curved space is considered -- and the absolute magnitude of SNIa $M_{abs}$ for the analyses that use SNIa data\footnote{In cases where a varying $\alpha(z)$ is assumed in the mZKDR equation, the free parameters of the analysis are detailed in Table \ref{tab:models}.}. We use uniform priors for all parameters, which are shown in Table \ref{Tab: prior1}. We sample our posterior distributions using the EMCEE python library \citep{emcee} \footnote{The sampling algorithm is Affine Invariant Markov Chain Monte Carlo Ensemble Sampler while the convergence criterion is based on the autocorrelation time. For post-processing of the chains and the confidence contours, we used the GetDist library \citep{2025JCAP...08..025L}.}.

Our first general comment, derived from the analysis of all tables and figures, is that the PPS dataset is much more constraining than either H0LiCOW or SH0ES+HF. The reason for this, lies in the amount of data in the PPS dataset, which contains one order of magnitude more data points than SH0ES+HF and two orders of magnitude more data points than H0LiCOW. Consequently, when these datasets are combined in joint analyses, the results are predominantly determined by the PPS contribution. Another point worth emphasizing is that our inferred uncertainties are larger than those reported by the H0LiCOW collaboration \citep{2020MNRAS.498.1420W}. This difference arises primarily from the choice of priors on both $\Omega_m$ and, in non-flat scenarios, $\Omega_k$. 
However, inspection of the chains in that analysis  reveals that the sampled values pile up against the imposed prior bounds. Since these priors are intended to be non-informative, the prior ranges should be broadened in such cases to ensure a reliable exploration of the parameter space and robust parameter inference. \\
Within the flat ZKDR framework, our analysis shows that the most constraining data set for $\alpha$ is PPS, while H0LiCOW provides weaker limits (see Table~\ref{Tab: nook} and Figures~\ref{fig:fig0} and \ref{fig:fig1}). Besides, the SH0ES+HF dataset fails to impose any meaningful constraints in this framework. Since all results are consistent with  $\alpha = 1$ ($\Lambda$CDM model), the most constraining datasets yield higher values for the lower limit of $\alpha$. Conversely, when examining the flat mZKDR framework, we find that none of the datasets considered here is capable of effectively limiting the value of $\alpha$. 
Regarding the matter density parameter $\Omega_m$, our results indicate that within the flat ZKDR framework, PPS imposes particularly stringent constraints on $\Omega_m$, while both SH0ES+HF and H0LiCOW provide less restrictive limits. In the mZKDR framework, our analysis shows that constraints derived from the H0LiCOW and SH0ES+HF datasets are comparable to those observed within the ZKDR framework. The PPS dataset, however, provides a more restrictive constraint within the mZKDR framework relative to the other datasets, though this constraint remains notably less stringent than its counterpart obtained within the ZKDR framework. We also note that the mZKDR distance relation indicates slightly higher values for $\Omega_m$ with respect to ZKDR and $\Lambda$CDM. These observed differences between models can be attributed to the parameter degeneracy between $\alpha$ and $\Omega_m$ in the mZKDR framework (as shown in Eq. \ref{ec:mDR} and discussed in Sect.\ref{theory}), a degeneracy that is broken in the  ZKDR model (see Eq. \ref{ec:DR}). Accordingly, the confidence contours in the $\alpha-\Omega_m$ plane in Figs. \ref{fig:sfig02} and \ref{fig:sfig2} show strong correlations for the mZKDR model while for the  ZKDR case only very weak degeneracies appear.
Regarding the Hubble constant $H_0$,  both SH0ES+HF and PPS datasets provide substantially more stringent constraints than H0LiCOW across all models examined. We also note that when using H0LiCOW data,  the estimated  $H_0$ values shift to lower values across all models. This effect is more pronounced for ZKDR, which exhibits a tension with the CMB inferred value at  the  1.1$\sigma$ level,  less than mZKDR ($\sim$2.1$\sigma$) and $\Lambda$CDM (2$\sigma$)\footnote{Here $N_\sigma$ is calculated using the 'rule of Thumb' which quantifies the disagreement between two inferred parameters $\mu_{A/B}$ with variance $\sigma_{A/B}$ in terms of $N_\sigma=\frac{\mu_A - \mu_B}{\sqrt{\sigma_A^2 + \sigma_B^2}.}$}.  However, when PPS data are incorporated, the tension increases to exceed $5\sigma$ for all models.  
On the other hand, only very weak correlations are  observed in the $H_0-\alpha$  plane across all data sets, which explains why the modified distance relations considered here  cannot solve or alleviate the $H_0$ tension. \\
We now extend our analysis to the non-flat case (see Table \ref{Tab: okfree} and Fig. \ref{fig:fig2}). We observe that $\alpha$ is only weakly constrained, with similar values obtained across all data sets both in the ZKDR and mZKDR frameworks.  Regarding the matter density parameter $\Omega_m$ the constraints for both frameworks remain comparable across all datasets, with those obtained from PPS being the most stringent ones. The difference between these results and those previously described for flat geometries can be attributed to the additional degree of freedom introduced in non-flat models, which consequently generates more parameter degeneracies within the theoretical framework. An interesting feature is that the parameter space inferred for the ZKDR model appears slightly more constrained than that obtained for $\Lambda$CDM when using only H0LiCOW data. However, this behavior may be driven 
by 
the correlation between $\alpha$ and $\Omega_k$\footnote{This fact may partially break the geometric degeneracies among $\Omega_m$, $\Omega_k$ and $H_0$, leading to a more localized likelihood and consequently narrower marginalized constraints on these parameters.
}, which is nearly absent in the mZKDR case (see the correlation matrices in \ref{app}), as well as by the limited amount of available observational data.
With regards to $H_0$, all  studied models yield higher values than those obtained for the flat case when using only H0LiCOW data.
The difference in the behaviour of the models, can be explained by the addition of an extra parameter ($\Omega_k$) which is also correlated with $H_0$ (see Fig. \ref{fig:sfig3}), but this correlation is not apparent when using the different SnIa datasets. In short, we find no evidence that the non-flat ZDKR models may alleviate the Hubble tension since when including PPS the discrepancy with the Planck value remains greater than 5$\sigma$ across all models.  
We now turn our attention to the mZKDR1 and mZKDR2 models (see Table \ref{tab:models} for details), which are shown in Table \ref{Tab: all mDRm}. Following the general trend, $\alpha_0$ (mZKDR1) is more tightly constrained by the PPS data set, and the same behavior is observed for $\beta_0$ (mZKDR2). At the same time, none of the data sets provide meaningful constraints for either $\alpha_1$ or $\gamma$. This can be explained, since these parameters quantify the time dependence of $\alpha$ which is more difficult to constrain. However,  our results remain consistent with a time-varying $\alpha$ at the $2\sigma$ level for both models. With regards to $\Omega_m$, the general trend observed in the other models remains, with the PPS data set providing the most stringent constraints. However,  the values of $\Omega_m$ estimated under the mZDKR1 model are lower than those obtained when assuming the mZKDR2 model. In contrast to the other models studied here, including $\Lambda$CDM, when considering the H0LiCOW data, the inferred values of $H_0$ exhibit a smaller shift toward higher values and, as a result, the disagreement with the CMB value is at the 2.5$\sigma$ level for mZKDR1 and 2.6$\sigma$ for mZKDR2, while when all data is used they are at more than 5$\sigma$.
In summary, our results do not provide strong evidence that the ZKDR and mZKDR models considered here can alleviate the Hubble tension. Even though  Eqs. \ref{ec:DR} and \ref{ec:mDR} do not depend on $H_0$, this parameter enters the calculation of $D_A$ as an inverse multiplicative factor. Therefore, some degeneracy in the $H_0$–$\alpha$ plane is  expected, but the statistical analyses performed in this paper show that this degeneracy is small (see Figs. \ref{fig:fig0}, \ref{fig:fig1} and \ref{fig:fig2}).
 Finally, we employ the AIC and BIC model comparison criteria to assess which models are favored by the data. The results presented in Table~\ref{tab:statistics} indicate that there is  weak evidence in favor of $\Lambda$CDM against the ZKDR model, whereas the remaining models considered in this work are strongly disfavored.

\begin{table*}[ht]
    \centering
    \footnotesize
    \caption{The estimates of different statistical criteria obtained for each analyzed approach.}
    \label{tab:statistics}
    \begin{tabular}{cccccccc}
        \hline
    \hline
       Model  & $\chi^2$ & $\chi_{\nu}^2$ & AIC & BIC  & $\Delta \chi^2$ & $\Delta$AIC & $\Delta$BIC\\
       \hline
        ZKDR & 1553.98 & 0.94 & 1561.98 & 1583.64 & -0.02&-2.02 & -7.43\\
       mZKDR& 1583.56 &0.95 &1591.56 &1613.22& -29.60& -31.63&-37.01  \\
       $\Lambda$CDM &1553.96& 0.94& 1559.96&  1576.21& --& --&--\\
        non-flat ZKDR &1583.70 & 0.96 &1593.70  &1620.78 &-29.74 &-33.74 &-44.57\\
       non-flat mZKDR& 1582.84  & 0.95& 1592.84&1619.92&-28.88 &-32.88 & -43.71\\
       non-flat $\Lambda$CDM & 1582.71&0.95 & 1590.71& 1612.38&-28.75 &-22.75 &-36.17\\
       mZKDR1&1583.59 & 0.96 & 1593.59&1620.67 & -29.63&-33.63 &-44.46\\
       mZKDR2 &1583.58 &0.96 &1593.58& 1620.66 &-29.62 &-33.62 &-44.45\\
   \hline
    \hline
    \end{tabular}
\end{table*}

\section{Conclusions}\label{conclusions}

In this study, we examine the impact of small-scale inhomogeneities on the propagation of light from distant sources, with particular emphasis on their implications for the Hubble tension. We employ the Zeldovich-Kantowski-Dyer-Roeder (ZKDR) approximation, along with a modified variant, to model these inhomogeneities. Our analysis encompasses both flat and curved cosmological models, allowing the smoothing parameter within the ZKDR distance relation to vary with redshift. To assess these scenarios, we use current observational data from the Pantheon$+$ compilation, as well as the SH0ES and H0LiCOW collaborations. 

Our main conclusion is that neither the ZKDR approximation nor its modification can solve or even alleviate the Hubble tension. This result is consistent with findings from previous studies. For instance, \cite{Odderskov_2016} investigated the effects of local inhomogeneities in the velocity field on the estimation of $H_0$ at low redshifts by computing the redshift-distance relationship for mock sources in N-body simulations, which are subsequently contrasted with results derived from the conventional methodology to estimate $H_0$. Moreover, \cite{Miura_2024} explored the inhomogeneities of the universe within the framework of Newtonian cosmology, using the adhesion model for collapsed regions that adhere to the Zeldovich approximation. Through this approach, the authors determine the luminosity distance and redshift of the source by transporting the wave vector along null geodesics, thereby making possible the estimation of $H_0$. 

Finally, we underscore that the tension surrounding the $H_0$ measurement remains one of the most pressing unresolved issues in cosmology, with the potential to uncover physics beyond the standard $\Lambda$CDM model. Among the various approaches to address this issue, we have explored a possibility that does not rely on introducing new physics, but only on the effects of small-scale inhomogeneities on light propagation. We believe that upcoming and ongoing surveys will provide higher-quality data, especially on time-delay lensing, allowing us to validate or contest the results and conclusions of this work.

\begin{acknowledgements}
      L.K. and S.L. are supported by grant PIP 11220200100729CO CONICET, grant 20020170100129BA UBACYT and grant SG002 UNLP. J.S.A is supported by Conselho Nacional de Desenvolvimento Científico e Tecnológico (CNPq) under Grant No. 307683/2022-2 and by Funda\c{c}\~ao de Amparo \`a Pesquisa do Estado do Rio de Janeiro (FAPERJ) under Grant No. 299312 (2023).
 
\end{acknowledgements}

\bibliographystyle{aa} 
\bibliography{example}

@article{DESI:2025fii,
      author = "Lodha, K. and others",
      title = "{Extended dark energy analysis using DESI DR2 BAO measurements}",
      journal = {PRD},
         year = 2025,
        month = oct,
       volume = {112},
       number = {8},
          eid = {083511},
        pages = {083511},
          doi = {10.1103/w4c6-1r5j},
archivePrefix = {arXiv},
       eprint = {2503.14743},
 primaryClass = {astro-ph.CO},
       adsurl = {https://ui.adsabs.harvard.edu/abs/2025PhRvD.112h3511L}
}

@article{Aiola_2020,
  author = {{Aiola et al.}},
  title = "{The Atacama Cosmology Telescope: DR4 maps and cosmological parameters}",
      journal = {\jcap},
         year = 2020,
        month = dec,
       volume = {2020},
       number = {12},
          eid = {047},
        pages = {047},
          doi = {10.1088/1475-7516/2020/12/047},
archivePrefix = {arXiv},
       eprint = {2007.07288},
 primaryClass = {astro-ph.CO},
       adsurl = {https://ui.adsabs.harvard.edu/abs/2020JCAP...12..047A}
}

@ARTICLE{2021ApJ...919...16F,
       author = {{Freedman}, Wendy L.},
        title = "{Measurements of the Hubble Constant: Tensions in Perspective}",
      journal = {\apj},
         year = 2021,
        month = sep,
       volume = {919},
       number = {1},
          eid = {16},
        pages = {16},
          doi = {10.3847/1538-4357/ac0e95},
archivePrefix = {arXiv},
       eprint = {2106.15656},
 primaryClass = {astro-ph.CO},
       adsurl = {https://ui.adsabs.harvard.edu/abs/2021ApJ...919...16F}
}

@article{DiValentino:2021izs,
    author = "Di Valentino, Eleonora and Mena, Olga and Pan, Supriya and Visinelli, Luca and Yang, Weiqiang and Melchiorri, Alessandro and Mota, David F. and Riess, Adam G. and Silk, Joseph",
    title = "{In the realm of the Hubble tension\textemdash{}a review of solutions}",
    eprint = "2103.01183",
    archivePrefix = "arXiv",
    primaryClass = "astro-ph.CO",
    reportNumber = "IPPP/20/108",
    doi = "10.1088/1361-6382/ac086d",
    journal = "CQG",
    volume = "38",
    number = "15",
    pages = "153001",
    year = "2021"
}

@article{Alcaniz:2019kah,
    author = "Alcaniz, Jailson and Bernal, Nicol\'as and Masiero, Antonio and Queiroz, Farinaldo S.",
    title = "{Light dark matter: A common solution to the lithium and H 0 problems}",
    eprint = "1912.05563",
    archivePrefix = "arXiv",
    primaryClass = "astro-ph.CO",
    reportNumber = "PI/UAN-2019-662FT,IIPDM-2019",
    doi = "10.1016/j.physletb.2020.136008",
    journal = "PLB",
    volume = "812",
    pages = "136008",
    year = "2021"
}

@ARTICLE{1965SvA.....8..854D,
       author = {{Dashevskii}, V.~M. and {Zel'dovich}, Ya. B.},
        title = "{Propagation of Light in a Nonhomogeneous Nonflat Universe II}",
      journal = {\sovast},
         year = 1965,
        month = jun,
       volume = {8},
        pages = {854},
       adsurl = {https://ui.adsabs.harvard.edu/abs/1965SvA.....8..854D}
}

@ARTICLE{1969ApJ...155...89K,
       author = {{Kantowski}, R.},
        title = "{Corrections in the Luminosity-Redshift Relations of the Homogeneous Fried-Mann Models}",
      journal = {\apj},
         year = 1969,
        month = jan,
       volume = {155},
        pages = {89},
          doi = {10.1086/149851},
       adsurl = {https://ui.adsabs.harvard.edu/abs/1969ApJ...155...89K}
}

@ARTICLE{1964SvA.....8...13Z,
       author = {{Zel'dovich}, Ya. B.},
        title = "{Observations in a Universe Homogeneous in the Mean}",
      journal = {\sovast},
         year = 1964,
        month = aug,
       volume = {8},
        pages = {13},
       adsurl = {https://ui.adsabs.harvard.edu/abs/1964SvA.....8...13Z}
}

@ARTICLE{2004IJMPD..13.1309A,
       author = {{Alcaniz}, Jailson S. and {Lima}, Jos{\'e} A.~S. and {Silva}, Raimundo},
        title = "{Mass Inhomogeneities and the Angular Size-Redshift Relation}",
      journal = {IJMPD},
         year = 2004,
        month = jan,
       volume = {13},
       number = {7},
        pages = {1309-1313},
          doi = {10.1142/S0218271804005468},
       adsurl = {https://ui.adsabs.harvard.edu/abs/2004IJMPD..13.1309A}
}

@article{Helbig:2019jcm,
    author = "Helbig, Phillip",
    title = "{Calculation of distances in cosmological models with small-scale inhomogeneities and their use in observational cosmology: a review}",
    eprint = "1912.12269",
    archivePrefix = "arXiv",
    primaryClass = "astro-ph.CO",
    doi = "10.21105/astro.1912.12269",
    journal = "OJAp",
    volume = "3",
    pages = "1",
    year = "2020"
}

@article{Efstathiou:2021ocp,
    author = "Efstathiou, George",
    title = "{To H0 or not to H0?}",
    eprint = "2103.08723",
    archivePrefix = "arXiv",
    primaryClass = "astro-ph.CO",
    doi = "10.1093/mnras/stab1588",
    journal = "MNRAS",
    volume = "505",
    number = "3",
    pages = "3866--3872",
    year = "2021"
}

@article{daCosta:2023mow,
    author = "da Costa, Simony Santos and da Silva, D\^eivid R. and de Jesus, \'Alvaro S. and Pinto-Neto, Nelson and Queiroz, Farinaldo S.",
    title = "{The H $_{0}$ trouble: confronting non-thermal dark matter and phantom cosmology with the CMB, BAO, and Type Ia supernovae data}",
    eprint = "2311.07420",
    archivePrefix = "arXiv",
    primaryClass = "astro-ph.CO",
    doi = "10.1088/1475-7516/2024/04/035",
    journal = "JCAP",
    volume = "04",
    pages = "035",
    year = "2024"
}

@article{Poulin:2018cxd,
    author = "Poulin, Vivian and Smith, Tristan L. and Karwal, Tanvi and Kamionkowski, Marc",
    title = "{Early Dark Energy Can Resolve The Hubble Tension}",
    eprint = "1811.04083",
    archivePrefix = "arXiv",
    primaryClass = "astro-ph.CO",
    doi = "10.1103/PhysRevLett.122.221301",
    journal = "PRL",
    volume = "122",
    number = "22",
    pages = "221301",
    year = "2019"
}

@article{Karwal:2016vyq,
    author = "Karwal, Tanvi and Kamionkowski, Marc",
    title = "{Dark energy at early times, the Hubble parameter, and the string axiverse}",
    eprint = "1608.01309",
    archivePrefix = "arXiv",
    primaryClass = "astro-ph.CO",
    doi = "10.1103/PhysRevD.94.103523",
    journal = "PRD",
    volume = "94",
    number = "10",
    pages = "103523",
    year = "2016"
}

@article{Alcaniz:2022oow,
    author = "Alcaniz, Jailson S. and Neto, Jacinto P. and Queiroz, Farinaldo S. and da Silva, Deivid R. and Silva, Raimindo",
    title = "{The Hubble constant troubled by dark matter in non-standard cosmologies}",
    eprint = "2211.14345",
    archivePrefix = "arXiv",
    primaryClass = "astro-ph.CO",
    doi = "10.1038/s41598-022-24608-5",
    journal = "SR",
    volume = "12",
    number = "1",
    pages = "20113",
    year = "2022",
    note = "[Erratum: SR, 13, 209 (2023)]"
}

@article{Efstathiou:2024dvn,
    author = "Efstathiou, George",
    title = "{Challenges to the \ensuremath{\Lambda}CDM cosmology}",
    eprint = "2406.12106",
    archivePrefix = "arXiv",
    primaryClass = "astro-ph.CO",
    doi = "10.1098/rsta.2024.0022",
    journal = "PTRA",
    volume = "383",
    number = "2290",
    pages = "20240022",
    year = "2025"
}

@ARTICLE{2020ApJ...891...57F,
       author = {{Freedman}, Wendy L. and {Madore}, Barry F. and {Hoyt}, Taylor and {Jang}, In Sung and {Beaton}, Rachael and {Lee}, Myung Gyoon and {Monson}, Andrew and {Neeley}, Jill and {Rich}, Jeffrey},
        title = "{Calibration of the Tip of the Red Giant Branch}",
      journal = {\apj},
         year = 2020,
        month = mar,
       volume = {891},
       number = {1},
          eid = {57},
        pages = {57},
          doi = {10.3847/1538-4357/ab7339},
archivePrefix = {arXiv},
       eprint = {2002.01550},
 primaryClass = {astro-ph.GA},
       adsurl = {https://ui.adsabs.harvard.edu/abs/2020ApJ...891...57F}
}

@article{PhysRevD.108.023510,
  title = {Measurement of the CMB temperature power spectrum and constraints on cosmology from the SPT-3G 2018 $TT$, $TE$, and $EE$ dataset},
  author = {{Balkenhol et al.}},
  collaboration = {SPT-3G Collaboration},
  journal = {PRD},
  volume = {108},
  issue = {2},
  pages = {023510},
  numpages = {34},
  year = {2023},
  month = {Jul},
  publisher = {American Physical Society},
  doi = {10.1103/PhysRevD.108.023510},
  url = {https://link.aps.org/doi/10.1103/PhysRevD.108.023510}
}

@article{Odderskov_2016,
title = {The local value of H0 in an inhomogeneous universe},
doi = {10.1088/1475-7516/2016/02/001},
url = {https://dx.doi.org/10.1088/1475-7516/2016/02/001},
year = {2016},
month = {feb},
publisher = {},
volume = {2016},
number = {02},
pages = {001},
author = {I. Odderskov and S.M. Koksbang and S. Hannestad},
journal = {JCAP}
}

@article{Miura_2024,
doi = {10.1088/1475-7516/2024/05/126},
url = {https://dx.doi.org/10.1088/1475-7516/2024/05/126},
year = {2024},
month = {may},
publisher = {IOP Publishing},
volume = {2024},
number = {05},
pages = {126},
author = {Taishi Miura and Takahiro Tanaka},
title = {Remarks on overestimating the effects of inhomogeneities on the Hubble constant},
journal = {JCAP}
}

@ARTICLE{2022ApJ...938..113S,
       author = {{Scolnic}, Dan and {Brout}, Dillon and {Carr}, Anthony and et al.},
        title = "{The Pantheon+ Analysis: The Full Data Set and Light-curve Release}",
      journal = {ApJ},
         year = 2022,
        month = oct,
       volume = {938},
       number = {2},
          eid = {113},
        pages = {113},
          doi = {10.3847/1538-4357/ac8b7a},
archivePrefix = {arXiv},
       eprint = {2112.03863},
 primaryClass = {astro-ph.CO},
       adsurl = {https://ui.adsabs.harvard.edu/abs/2022ApJ...938..113S}
}

@ARTICLE{2022ApJ...938..110B,
       author = {{Brout}, Dillon and {Scolnic}, Dan and et al.},
        title = "{The Pantheon+ Analysis: Cosmological Constraints}",
      journal = {ApJ},
         year = 2022,
        month = oct,
       volume = {938},
       number = {2},
          eid = {110},
        pages = {110},
          doi = {10.3847/1538-4357/ac8e04},
archivePrefix = {arXiv},
       eprint = {2202.04077},
 primaryClass = {astro-ph.CO},
       adsurl = {https://ui.adsabs.harvard.edu/abs/2022ApJ...938..110B}
}

@ARTICLE{Tripp1998,
       author = {{Tripp}, Robert},
        title = "{A two-parameter luminosity correction for Type IA supernovae}",
      journal = {A$\&$A},
         year = 1998,
        month = mar,
       volume = {331},
        pages = {815-820},
       adsurl = {https://ui.adsabs.harvard.edu/abs/1998A&A...331..815T}
}

@ARTICLE{Leizerovich2022,
       author = {{Leizerovich}, Mat{\'\i}as and {Kraiselburd}, Lucila and {Landau}, Susana and {Sc{\'o}ccola}, Claudia G.},
        title = "{Testing f (R ) gravity models with quasar x-ray and UV fluxes}",
      journal = {PRD},
         year = 2022,
        month = may,
       volume = {105},
       number = {10},
          eid = {103526},
        pages = {103526},
          doi = {10.1103/PhysRevD.105.103526},
archivePrefix = {arXiv},
       eprint = {2112.01492},
 primaryClass = {astro-ph.CO},
       adsurl = {https://ui.adsabs.harvard.edu/abs/2022PhRvD.105j3526L}
}

@ARTICLE{Negrelli2020,
       author = {{Negrelli}, Carolina and {Kraiselburd}, Lucila and {Landau}, Susana and {Sc{\'o}ccola}, Claudia G.},
        title = "{Testing Modified Gravity theory (MOG) with Type Ia Supernovae, Cosmic Chronometers and Baryon Acoustic Oscillations}",
      journal = {JCAP},
         year = 2020,
        month = jul,
       volume = {2020},
       number = {7},
          eid = {015},
        pages = {015},
          doi = {10.1088/1475-7516/2020/07/015},
archivePrefix = {arXiv},
       eprint = {2004.13648},
 primaryClass = {astro-ph.CO},
       adsurl = {https://ui.adsabs.harvard.edu/abs/2020JCAP...07..015N}
}

@ARTICLE{2012MNRAS.426.1121C,
       author = {{Clarkson}, Chris and {Ellis}, George F.~R. and {Faltenbacher}, Andreas and {Maartens}, Roy and {Umeh}, Obinna and {Uzan}, Jean-Philippe},
        title = "{(Mis)interpreting supernovae observations in a lumpy universe}",
      journal = {MNRAS},
         year = 2012,
        month = oct,
       volume = {426},
       number = {2},
        pages = {1121-1136},
          doi = {10.1111/j.1365-2966.2012.21750.x},
archivePrefix = {arXiv},
       eprint = {1109.2484},
 primaryClass = {astro-ph.CO},
       adsurl = {https://ui.adsabs.harvard.edu/abs/2012MNRAS.426.1121C}
}

@ARTICLE{2006astro.ph..9129S,
       author = {{Santos}, R.~C. and {Lima}, J.~A.~S.},
        title = "{ZKDR Distance, Angular Size and Phantom Cosmology}",
        journal = {arXiv e-prints},
    eprint = "0609129",
    archivePrefix = "arXiv",
    primaryClass = "astro-ph.CO",
    month = "9",
    year = "2006"
}

@ARTICLE{2011MNRAS.412.1937B,
       author = {{Bolejko}, K.},
        title = "{Weak lensing and the Dyer-Roeder approximation}",
      journal = {MNRAS},
         year = 2011,
        month = apr,
       volume = {412},
       number = {3},
        pages = {1937-1942},
          doi = {10.1111/j.1365-2966.2010.18031.x},
archivePrefix = {arXiv},
       eprint = {1011.3876},
 primaryClass = {astro-ph.CO},
       adsurl = {https://ui.adsabs.harvard.edu/abs/2011MNRAS.412.1937B}
}

@ARTICLE{2021MNRAS.503.3179K,
       author = {{Kalomenopoulos}, Marios and {Khochfar}, Sadegh and {Gair}, Jonathan and {Arai}, Shun},
        title = "{Mapping the inhomogeneous Universe with standard sirens: degeneracy between inhomogeneity and modified gravity theories}",
      journal = {MNRAS},
         year = 2021,
        month = may,
       volume = {503},
       number = {3},
        pages = {3179-3193},
          doi = {10.1093/mnras/stab557},
archivePrefix = {arXiv},
       eprint = {2007.15020},
 primaryClass = {astro-ph.CO},
       adsurl = {https://ui.adsabs.harvard.edu/abs/2021MNRAS.503.3179K}
}

@BOOK{1992grle.book.....S,
       author = {{Schneider}, Peter and {Ehlers}, J{\"u}rgen and {Falco}, Emilio E.},
        title = "{Gravitational Lenses}",
         year = 1992,
          doi = {10.1007/978-3-662-03758-4},
        publisher= {Springer},
       adsurl = {https://ui.adsabs.harvard.edu/abs/1992grle.book.....S}
}

@ARTICLE{2024arXiv240811031P,
 title = {Hubble tension or distance ladder crisis?},
  author = {Perivolaropoulos, Leandros},
  journal = {PRD},
  volume = {110},
  issue = {12},
  pages = {123518},
  numpages = {45},
  year = {2024},
  month = {Dec},
  publisher = {American Physical Society},
  doi = {10.1103/PhysRevD.110.123518},
  url = {https://link.aps.org/doi/10.1103/PhysRevD.110.123518}
}

@ARTICLE{2024arXiv240806153F,
author = "Freedman, Wendy L. and Madore, Barry F. and Hoyt, Taylor J. and Jang, In Sung and Lee, Abigail J. and Owens, Kayla A.",
    title = "{Status Report on the Chicago-Carnegie Hubble Program (CCHP): Measurement of the Hubble Constant Using the Hubble and James Webb Space Telescopes}",
    eprint = "2408.06153",
    archivePrefix = "arXiv",
    primaryClass = "astro-ph.CO",
    doi = "10.3847/1538-4357/adce78",
    journal = "ApJ",
    volume = "985",
    number = "2",
    pages = "203",
    year = "2025",
    note = "[Erratum: ApJ, 993, 252 (2025)]"
}

@article{Sousa-Neto:2025gpj,
    author = "Sousa-Neto, Agripino and Bengaly, Carlos and Gonz\'alez, Javier E. and Alcaniz, Jailson",
    title = "{No evidence for dynamical dark energy from DESI and SN data: a symbolic regression analysis}",
    journal = {arXiv e-prints},
    eprint = "2502.10506",
    archivePrefix = "arXiv",
    primaryClass = "astro-ph.CO",
    month = "2",
    year = "2025"
}

@article{Dinda:2024ktd,
    author = "Dinda, Bikash R. and Maartens, Roy",
    title = "{Model-agnostic assessment of dark energy after DESI DR1 BAO}",
    eprint = "2407.17252",
    archivePrefix = "arXiv",
    primaryClass = "astro-ph.CO",
    doi = "10.1088/1475-7516/2025/01/120",
    journal = "JCAP",
    volume = "01",
    pages = "120",
    year = "2025"
}

@article{DESI:2024mwx,
    author = "Adame, A. G. and others",
    collaboration = "DESI",
    title = "{DESI 2024 VI: cosmological constraints from the measurements of baryon acoustic oscillations}",
    eprint = "2404.03002",
    archivePrefix = "arXiv",
    primaryClass = "astro-ph.CO",
    reportNumber = "FERMILAB-PUB-24-0154-PPD",
    doi = "10.1088/1475-7516/2025/02/021",
    journal = "JCAP",
    volume = "02",
    pages = "021",
    year = "2025"
}

@article{Efstathiou:2024xcq,
    author = {Efstathiou, George},
    title = {Evolving dark energy or supernovae systematics?},
    journal = {MNRAS},
    volume = {538},
    number = {2},
    pages = {875-882},
    year = {2025},
    month = {02},
    issn = {0035-8711},
    doi = {10.1093/mnras/staf301},
    url = {https://doi.org/10.1093/mnras/staf301},
    eprint = {https://academic.oup.com/mnras/article-pdf/538/2/875/62200599/staf301.pdf},
}

@article{DESI:2025zgx,
    author = "Abdul Karim, M. and others and DESI Collaboration",
        title = "{DESI DR2 results. II. Measurements of baryon acoustic oscillations and cosmological constraints}",
      journal = {PRD},
         year = 2025,
        month = oct,
       volume = {112},
       number = {8},
          eid = {083515},
        pages = {083515},
          doi = {10.1103/tr6y-kpc6},
archivePrefix = {arXiv},
       eprint = {2503.14738},
 primaryClass = {astro-ph.CO},
       adsurl = {https://ui.adsabs.harvard.edu/abs/2025PhRvD.112h3515A}
}

@ARTICLE{Aghanim:2018eyx,
       author = {{Planck Collaboration: Aghanim}, N. and {et al.}},
        title = "{Planck 2018 results. VI. Cosmological parameters}",
      journal = {A$\&$A},
         year = 2020,
        month = sep,
       volume = {641},
          eid = {A6},
        pages = {A6},
          doi = {10.1051/0004-6361/201833910},
archivePrefix = {arXiv},
       eprint = {1807.06209},
 primaryClass = {astro-ph.CO},
       adsurl = {https://ui.adsabs.harvard.edu/abs/2020A&A...641A...6P}
}

@ARTICLE{2024JCAP...04..059K,
       author = {{Khalife}, Ali Rida and {Zanjani}, Maryam Bahrami and {Galli}, Silvia and {G{\"u}nther}, Sven and {Lesgourgues}, Julien and {Benabed}, Karim},
        title = "{Review of Hubble tension solutions with new SH0ES and SPT-3G data}",
      journal = {JCAP},
         year = 2024,
        month = apr,
       volume = {2024},
       number = {4},
          eid = {059},
        pages = {059},
          doi = {10.1088/1475-7516/2024/04/059},
archivePrefix = {arXiv},
       eprint = {2312.09814},
 primaryClass = {astro-ph.CO},
       adsurl = {https://ui.adsabs.harvard.edu/abs/2024JCAP...04..059K}
}

@ARTICLE{2020MNRAS.498.1420W,
       author = {{Wong}, Kenneth C. and {Suyu}, Sherry H. and {Chen}, Geoff C. -F. and {Rusu}, Cristian E. and et al.},
        title = "{H0LiCOW - XIII. A 2.4 per cent measurement of H$_{0}$ from lensed quasars: 5.3{\ensuremath{\sigma}} tension between early- and late-Universe probes}",
      journal = {MNRAS},
         year = 2020,
        month = oct,
       volume = {498},
       number = {1},
        pages = {1420-1439},
          doi = {10.1093/mnras/stz3094},
archivePrefix = {arXiv},
       eprint = {1907.04869},
 primaryClass = {astro-ph.CO},
       adsurl = {https://ui.adsabs.harvard.edu/abs/2020MNRAS.498.1420W}
}

@article{Refsdal:1964nw,
    author = "Refsdal, S.",
    title = "{On the possibility of determining Hubble's parameter and the masses of galaxies from the gravitational lens effect}",
    journal = "MNRAS",
    volume = "128",
    pages = "307",
    year = "1964"
}

@article{Suyu:2009by,
    author = "Suyu, S. H. and Marshall, P. J. and Auger, M. W. and Hilbert, S. and Blandford, R. D. and Koopmans, L. V. E. and Fassnacht, C. D. and Treu, T.",
    title = "{Dissecting the Gravitational Lens B1608+656. II. Precision Measurements of the Hubble Constant, Spatial Curvature, and the Dark Energy Equation of State}",
    eprint = "0910.2773",
    archivePrefix = "arXiv",
    primaryClass = "astro-ph.CO",
    reportNumber = "SLAC-PUB-13811",
    doi = "10.1088/0004-637X/711/1/201",
    journal = "ApJ",
    volume = "711",
    pages = "201--221",
    year = "2010"
}

@ARTICLE{2022ApJ...934L...7R,
       author = {{Riess}, Adam G. and {Yuan}, Wenlong and {Macri}, Lucas M. and {Scolnic}, Dan and {Brout}, Dillon and {Casertano}, Stefano and {Jones}, David O. and {Murakami}, Yukei and {Anand}, Gagandeep S. and {Breuval}, Louise and {Brink}, Thomas G. and {Filippenko}, Alexei V. and {Hoffmann}, Samantha and {Jha}, Saurabh W. and {D'arcy Kenworthy}, W. and {Mackenty}, John and {Stahl}, Benjamin E. and {Zheng}, WeiKang},
        title = "{A Comprehensive Measurement of the Local Value of the Hubble Constant with 1 km s$^{-1}$ Mpc$^{-1}$ Uncertainty from the Hubble Space Telescope and the SH0ES Team}",
      journal = {\apjl},
         year = 2022,
        month = jul,
       volume = {934},
       number = {1},
          eid = {L7},
        pages = {L7},
          doi = {10.3847/2041-8213/ac5c5b},
archivePrefix = {arXiv},
       eprint = {2112.04510},
 primaryClass = {astro-ph.CO},
       adsurl = {https://ui.adsabs.harvard.edu/abs/2022ApJ...934L...7R}
}

@ARTICLE{2025arXiv250500659C,
       author = {{Chen}, Shi-Fan and {Zaldarriaga}, Matias},
        title = "{It's all Ok: curvature in light of BAO from DESI DR2}",
      journal = {\jcap},
         year = 2025,
        month = aug,
       volume = {2025},
       number = {8},
          eid = {014},
        pages = {014},
          doi = {10.1088/1475-7516/2025/08/014},
archivePrefix = {arXiv},
       eprint = {2505.00659},
 primaryClass = {astro-ph.CO},
       adsurl = {https://ui.adsabs.harvard.edu/abs/2025JCAP...08..014C}
}

@ARTICLE{2025arXiv250415190D,
author = {{Dinda}, Bikash R. and {Maartens}, Roy},
        title = "{Physical versus phantom dark energy after DESI: thawing quintessence in a curved background}",
      journal = {\mnras},
         year = 2025,
        month = sep,
       volume = {542},
       number = {1},
        pages = {L31-L35},
          doi = {10.1093/mnrasl/slaf063},
archivePrefix = {arXiv},
       eprint = {2504.15190},
 primaryClass = {astro-ph.CO},
       adsurl = {https://ui.adsabs.harvard.edu/abs/2025MNRAS.542L..31D}
}

@article{emcee,
  author = {{Foreman-Mackey}, D. and {Hogg}, D.~W. and {Lang}, D. and {Goodman}, J. },
  title = {emcee: The MCMC Hammer},
  journal = {PASP},
  year = 2013,
  volume = 125,
  pages = {306-312},
  eprint = {1202.3665},
  doi = {10.1086/670067}
}

@ARTICLE{2025JCAP...08..025L,
       author = {{Lewis}, Antony},
        title = "{GetDist: a Python package for analysing Monte Carlo samples}",
      journal = {\jcap},
         year = 2025,
        month = aug,
       volume = {2025},
       number = {8},
          eid = {025},
        pages = {025},
          doi = {10.1088/1475-7516/2025/08/025},
archivePrefix = {arXiv},
       eprint = {1910.13970},
 primaryClass = {astro-ph.IM},
       adsurl = {https://ui.adsabs.harvard.edu/abs/2025JCAP...08..025L}
}

@ARTICLE{2024JHEAp..41...30D,
       author = {{Dainotti}, M.~G. and {Bargiacchi}, G. and {Bogdan}, M. and {Capozziello}, S. and {Nagataki}, S.},
        title = "{On the statistical assumption on the distance moduli of Supernovae Ia and its impact on the determination of cosmological parameters}",
      journal = {JHEAp},
         year = 2024,
        month = mar,
       volume = {41},
        pages = {30-41},
          doi = {10.1016/j.jheap.2024.01.001},
       adsurl = {https://ui.adsabs.harvard.edu/abs/2024JHEAp..41...30D}
}

@ARTICLE{2025MNRAS.536..234L,
       author = {{Lovick}, Toby and {Dhawan}, Suhail and {Handley}, Will},
        title = "{Non-Gaussian likelihoods for Type Ia supernova cosmology: implications for dark energy and H$_{0}$}",
      journal = {\mnras},
         year = 2025,
        month = jan,
       volume = {536},
       number = {1},
        pages = {234-246},
          doi = {10.1093/mnras/stae2617},
archivePrefix = {arXiv},
       eprint = {2312.02075},
 primaryClass = {astro-ph.CO},
       adsurl = {https://ui.adsabs.harvard.edu/abs/2025MNRAS.536..234L}
}

\clearpage
\begin{appendix}
\clearpage
\onecolumn
\section{Results tables}\label{RT}

The subsequent tables present the results obtained from the different statistical analyzes performed in this study. 
\begin{table*}[!ht]
\centering
\footnotesize
\caption{Results from our statistical analysis for the ZKDR and mZKDR approximations and (FLRW) $\Lambda$CDM with $\Omega_k=0$ and $\alpha$ constant.}
\label{Tab: nook}
\begin{tabular}{l c c c c c}
\hline
Model & Data & $\alpha$       & $\Omega_m$          & $H_0$                       & $M_{abs}$      \\
\hline 
ZKDR & H0LiCOW                                              &   $0.723_{-0.060(0.491)}^{+0.277(0.277)} $    &  $ 0.387_{-0.271(0.337)}^{+0.121(0.423)}$   &  $ 71.153_{-1.873(5.534)}^{+3.286(4.942)} $      &  ---                  \\

& PPS    &    $ 0.750_{-0.072(0.311)}^{+0.250(0.250)} $    &  $ 0.360_{-0.028(0.050)}^{+0.026(0.054)} $   &  $73.480_{-0.984(1.940)}^{+0.986(1.987)} $      &  $-19.242_{-0.028(0.058)}^{+0.028(0.055)}$                  \\

& SH0ES+HF   &    $0.526_{-0.175(0.472)}^{+0.455(0.464)} $    &  $0.410_{-0.199(0.322)}^{+0.136(0.310)}  $   &  $73.225_{-1.307(2.154)}^{+0.946(2.396)}  $      &  $-19.247_{-0.029(0.059)}^{+0.030(0.059)}$                  \\

& H0LiCOW+  PPS      &    $0.845_{-0.040(0.218)}^{+0.153(0.155)}  $   &  $0.349_{-0.023(0.042)}^{+0.020(0.045)} $   &  $73.528_{-0.863(1.818)}^{+0.897(1.653)} $      &  $-19.242_{-0.027(0.050)}^{+0.025(0.051)} $                  \\

\hline \hline

& & $\alpha$       & $\Omega_m$          & $H_0$                       & $M_{abs}$      \\
\hline 
mZKDR & H0LiCOW                                              &   $0.523_{-0.317(0.434)}^{+0.312(0.477)} $    &  $ 0.525_{-0.234(0.394)}^{+0.230(0.413)} $   &  $ 72.717_{-1.698(4.297)}^{+2.485(3.971)} $      &  ---                  \\
& PPS       &    $0.630_{-0.223(0.360)}^{+0.258(0.370)}$   &  $0.457_{-0.122(0.145)}^{+0.050(0.196)} $   &  $73.710_{-1.118(2.115)}^{+0.932(1.919)}    $      &  $-19.239_{-0.029(0.057)}^{+0.031(0.057)} $                  \\

& SH0ES+HF      &    $0.535_{-0.255(0.421)}^{+0.358(0.464)}   $   &  $ 0.509_{-0.198(0.374)}^{+0.228(0.377)} $   &  $73.483_{-1.144(2.223)}^{+1.089(2.163)} $      &  $-19.245_{-0.028(0.058)}^{+0.030(0.055)}  $                  \\
& H0LiCOW+  PPS       &    $ 0.635_{-0.118(0.430)}^{+0.364(0.365)}   $   &  $0.463_{-0.139(0.158)}^{+0.035(0.245)} $   &  $ 73.401_{-0.880(1.792)}^{+0.896(1.754)}  $      &  $-19.249_{-0.026(0.054)}^{+0.026(0.050)}  $                  \\
\hline \hline

& &       & $\Omega_m$          & $H_0$                       & $M_{abs}$      \\
\hline 
$\Lambda$CDM & H0LiCOW                                              &   ---   &  $0.403_{-0.273(0.353)}^{+0.108(0.414)} $   &  $72.516_{-1.823(4.471)}^{+2.504(4.127)}  $      &  ---                  \\

& PPS    &  ---    &  $0.332_{-0.018(0.034)}^{+0.017(0.036)}  $  &  $73.594_{-0.922(2.036)}^{+1.101(2.049)}$      &  $-19.243_{-0.029(0.059)}^{+0.029(0.060)} $                  \\
& SH0ES+HF    &  ---    &  $0.394_{-0.155(0.287)}^{+0.143(0.295)}$  &  $73.195_{-1.103(2.160)}^{+1.108(2.195)} $      &  $-19.248_{-0.029(0.055)}^{+0.029(0.059)} $                  \\
& H0LiCOW+  PPS       &   ---    &  $0.336_{-0.018(0.033)}^{+0.018(0.036)} $   &  $73.685_{-0.916(1.809)}^{+0.891(1.625)}  $      &  $ -19.239_{-0.023(0.050)}^{+0.028(0.050)} $                  \\
\hline \hline
\end{tabular}
%}
\tablefoot{ Data used in these analysis are from H0LiCOW gravitational lenses, luminosity distances reported by Pantheon$^+$+SH0ES collaboration (PPS), and the ones employed by SH0ES. For each parameter, we present the mean value and the 68\% (95\%) confidence levels, or the upper limits obtained.}
\end{table*}

\begin{table*}[!ht]
\footnotesize
\centering
\caption{The same as in the previous table for non-flat geometries and assuming $\alpha$ constant.}
\label{Tab: okfree}
\begin{tabular}{l c c c c c c}
\hline
Model& Data & $\alpha$       & $\Omega_m$    &  $\Omega_k$      & $H_0$                       & $M_{abs}$      \\
\hline 
non-flat ZKDR & H0LiCOW                                              &   $0.586_{-0.143(0.502)}^{+0.414(0.414)}  $    &  $ 0.213_{-0.163(0.163)}^{+0.026(0.313)} $   & 
$0.454_{-0.177(0.481)}^{+0.312(0.448)}   $   &
$74.330_{-1.722(5.637)}^{+3.353(5.097)}   $      &  ---                  \\

&PPS    &    $ 0.624_{-0.219(0.341)}^{+0.237(0.376)} $    &  $ 0.440_{-0.130(0.202)}^{+0.111(0.200)} $   &  
$-0.141_{-0.235(0.359)}^{+0.196(0.343)}  $     &  
$73.465_{-0.982(2.002)}^{+0.953(1.895)} $      &  $-19.244_{-0.030(0.054)}^{+0.025(0.057)} $                  \\

&SH0ES+HF   &    $0.491_{-0.427(0.478)}^{+0.222(0.462)} $    &  $ 0.403_{-0.326(0.351)}^{+0.133(0.370)} $   &  
$0.032_{-0.530(0.530)}^{+0.177(0.599)}  $     &  
$73.142_{-1.267(2.142)}^{+1.019(2.247)}$      &  $-19.248_{-0.028(0.056)}^{+0.031(0.061)} $                  \\
& H0LiCOW+  PPS       &    $ 0.726_{-0.091(0.286)}^{+0.274(0.274)}  $   &  $ 0.438_{-0.113(0.152)}^{+0.083(0.179)}  $   & 
$-0.171_{-0.156(0.329)}^{+0.202(0.264)}  $      &
$73.396_{-0.913(1.770)}^{+0.881(1.839)}  $      &  $-19.249_{-0.025(0.049)}^{+0.025(0.051)}  $                  \\

\hline \hline
& & $\alpha$       & $\Omega_m$    &  $\Omega_k$      & $H_0$                       & $M_{abs}$      \\
\hline 
non-flat mZKDR & H0LiCOW                                              &   $  0.513_{-0.240(0.452)}^{+0.387(0.479)} $    &  $ 0.354_{-0.304(0.304)}^{+0.084(0.475)} $   & 
$0.320_{-0.234(0.683)}^{+0.458(0.567)} $   &
$74.405_{-1.886(4.602)}^{+2.872(4.336)}   $      &  ---                  \\

& PPS    &    $0.610_{-0.117(0.411)}^{+0.387(0.390)}   $    &  $0.429_{-0.149(0.201)}^{+0.074(0.262)} $   &  
$ 0.079_{-0.106(0.186)}^{+0.096(0.222)}  $     &  
$73.495_{-0.953(2.065)}^{+1.148(1.988)}  $      &  $-19.242_{-0.030(0.055)}^{+0.030(0.061)}  $                  \\

& SH0ES+HF    &    $0.517_{-0.276(0.453)}^{+0.345(0.477)}   $    &  $ 0.445_{-0.395(0.395)}^{+0.138(0.436)}  $   &  
$ 0.148_{-0.424(0.620)}^{+0.325(0.580)}  $     &  
$73.154_{-1.249(2.025)}^{+0.950(2.280)}   $      &  $-19.247_{-0.027(0.059)}^{+0.030(0.056)}  $                  \\
& H0LiCOW+  PPS       &    $0.556_{-0.170(0.468)}^{+0.405(0.403)}  $   &  $0.462_{-0.196(0.218)}^{+0.069(0.398)}  $   & 
$0.081_{-0.116(0.191)}^{+0.088(0.221)}  $      &
$73.486_{-0.885(1.633)}^{+1.089(1.857)} $      &  $ -19.243_{-0.029(0.053)}^{+0.026(0.051)}   $                  \\
\hline \hline

& & $\alpha$       & $\Omega_m$    &  $\Omega_k$      & $H_0$                       & $M_{abs}$      \\
\hline 
non-flat $\Lambda$CDM & H0LiCOW                                              &   ---    &  $ 0.259_{-0.208(0.209)}^{+0.040(0.392)}   $   & 
$ 0.364_{-0.234(0.571)}^{+0.378(0.537)} $   &
$ 74.559_{-2.055(5.291)}^{+2.837(4.501)}   $      &  ---                  \\

& PPS    &  ---    &  $ 0.300_{-0.060(0.101)}^{+0.049(0.109)}  $   &  
$0.088_{-0.127(0.235)}^{+0.134(0.265)}  $     &  
$73.459_{-1.056(1.923)}^{+0.953(2.024)} $      &  $-19.243_{-0.029(0.057)}^{+0.029(0.056)}  $                  \\

& SH0ES+HF    &  ---    & $ 0.398_{-0.289(0.348)}^{+0.170(0.367)}  $   & 
$0.028_{-0.524(0.528)}^{+0.156(0.597)} $      &
$73.215_{-1.157(2.090)}^{+1.100(2.346)}$      &  $ -19.246_{-0.031(0.060)}^{+0.029(0.057)} $ 
                 \\
& H0LiCOW+  PPS       &   ---   &  $ 0.290_{-0.047(0.096)}^{+0.049(0.096)}   $   & 
$ 0.111_{-0.130(0.214)}^{+0.098(0.236)} $      &
$73.643_{-0.908(1.735)}^{+0.866(1.761)} $      &  $ -19.237_{-0.026(0.055)}^{+0.026(0.049)} $                  \\
\hline \hline
\end{tabular}

\end{table*}

\begin{table*}[!ht]
\centering

\footnotesize
\caption{The same as in the previous table for flat geometries and assuming $\alpha$ as function of $z$.}
\label{Tab: all mDRm}
\begin{tabular}{l c c c c c c }
\hline
Model & Data & $\alpha_0$       &  $\alpha_1$       &$\Omega_m$          & $H_0$                       & $M_{abs}$      \\
\hline
mZKDR1& H0LiCOW                                              &   $< 0.558(0.914)$ &  $  0.541_{-0.160(0.472)}^{+0.456(0.457)}  $    &  $  0.461_{-0.248(0.362)}^{+0.195(0.401)} $   &  $  74.566_{-1.924(4.842)}^{+2.866(4.612)}$      &  ---                  \\

&H0LiCOW+  PPS       &    $  0.619_{-0.107(0.496)}^{+0.380(0.380)}   $   &  $ < 0.520(0.905)  $  &  $ 0.449_{-0.147(0.156)}^{+0.024(0.353)} $  &  $  73.869_{-0.949(1.645)}^{+0.760(1.850)} $      &  $ -19.237_{-0.023(0.049)}^{+0.027(0.051)}  $                  \\
\hline \hline

& & $\beta_0$       &  $\gamma$       &$\Omega_m$          & $H_0$                       & $M_{abs}$      \\
\hline 
mZKDR2 & H0LiCOW                                              &   $0.493_{-0.382(0.461)}^{+0.273(0.460)}   $ &  $  0.559_{-0.153(0.476)}^{+0.440(0.440)}  $    &  $  0.537_{-0.211(0.375)}^{+0.221(0.387)} $   &  $ 74.101_{-1.806(4.623)}^{+2.535(4.376)}   $      &  ---                  \\

& H0LiCOW+  PPS       &    $  0.592_{-0.176(0.443)}^{+0.408(0.407)}$   &  $ < 0.471(0.859)$   &$ 0.573_{-0.124(0.145)}^{+0.043(0.202)}  $   &  $ 73.869_{-0.927(1.750)}^{+0.823(1.760)} $      &  $ -19.238_{-0.023(0.054)}^{+0.027(0.047)} $                  \\
\hline \hline
\end{tabular}

\end{table*}
\clearpage
\twocolumn
\section{Correlation matrices}\label{app}

The following matrices present the parameter correlations for the non-flat ZKDR and non-flat mZKDR models, using exclusively H0LiCOW data. The parameters are ordered as $\alpha$, $\Omega_m$, $\Omega_k$ and $H_0$. In the case of the non-flat ZKDR model (upper one), a mild correlation between $\alpha$ and $\Omega_k$
 is observed, whereas this correlation is negligible in the mZKDR model.
\[
\begin{bmatrix}

 1.      &    0.1244055 & -0.23682401 & 0.06672599 \\
 0.1244055 &  1.     &    -0.5107811  & -0.72096984 \\
 -0.23682401  & -0.5107811 &  1.  &        0.51146289\\
 0.06672599 & -0.72096984  & 0.51146289  & 1.           
\end{bmatrix}
\]

\[
\begin{bmatrix}
 1.     &    -0.14047518 &  0.04698075 & -0.04457701\\
 -0.14047518 &  1.      &   -0.7764577 & -0.38876469\\
  0.04698075  & -0.7764577  &  1.  &        0.29348115\\
 -0.04457701 & -0.38876469  & 0.29348115 & 1.            
\end{bmatrix} 
\]
\end{appendix}
\end{document}